# The Structure of Dark Matter Halos

Giuseppe Tormen

*I.o.A., Univ. of Cambridge, Madingley Rd, Cambridge CB3 0HA, UK;
I.A.P., 98bis Boulevard Arago, 75014 Paris, FRANCE; current address:
M.P.A., 85740 Garching, GERMANY*

**Abstract.** I use $N$-body simulations to investigate the morphology and the dynamical evolution and properties of dark matter halos in clusters of galaxies. My sample consists of nine massive halos, coming from an Einstein-De Sitter universe with scale free power spectrum and spectral index $n = -1$. Halos are resolved by 20,000 particles each, on average, and have a dynamical resolution of 20-25 kpc, as shown by extensive tests. I find that the *average* density profiles of the halos are fitted by the Navarro, Frenk and White (1995) analytical fit with a root mean square (rms) residual of 20% within the Virial radius. The Hernquist (1990) analytical density profiles fits the same halos with an rms residual of 30%. The projected mass profiles of the simulated halos are in very good agreement with the profiles of rich galaxy clusters derived from strong and weak gravitational lensing observations.

## 1. Introduction

In this contribution I present some results from $N$-body simulations of the formation of galaxy clusters. In current cosmological models the mass and dynamics of galactic and supergalactic structures are dominated by some kind of non-baryonic, *dark* matter, which interacts with ordinary baryonic matter only through gravity. Therefore, in studies focussing on the dynamics of galaxy clusters, these can be regarded as halos made of collisionless dark matter. From the theoretical point of view one can study the structure of dark matter halos both analytically and numerically. The analytical work done so far is based on the *secondary infall* paradigma, (Gunn and Gott 1972), and calculations predicts that the density profile of the virialized halo should go as $\rho(r) \propto r^{-9/4}$. Hoffman and Shaham (1985) applied this paradigm to the gravitational instability theory of hierarchical clustering. In their calculations they assumed a Gaussian random field of initial density perturbation, and scale free power spectra $P(k) \propto k^n$. They found that the virialized structures originating from density peaks have density profiles whose shape depends on the spectral index $n$ as $\rho(r) \propto r^{-(9+3n)/(4+n)}$.

In reality the collapse of an initial overdensity is not so simple. In particular, motions are not purely spherical, and accretion does not happen in spherical shells (as assumed in the secondary infall model), but by aggregation of subclumps of matter already virialized. It is therefore very important to com-plement and compare the analytical studies with numerical simulations, which are not bound by such restrictions. However, due also to computational limits in present day machines, simulations of galaxy clusters have not yet reached a resolution sufficient to resolve the dark matter structure of the central $\approx 100$ kpc in a way free of numerical limitations.

The distribution of dark matter in the central region of the halos is of particular interest; for example, the ability of a cluster to act as a gravitational lens, producing multiple magnified images of background galaxies in the line of sight, crucially depends on the mass content of the very central part (few tens of kpc) of the cluster, hence on the slope of the density profile at that scale. The purpose of the present study is to show that such resolution can in fact be achieved with the current computational resources. Unveiling the structure of dark matter halos down to few tens of kpc from the halo center can provide important hindsight on the dynamics and the formation process of galaxy clusters, as well as give us more tools for the comparison of models with observations.

Section 2 presents the simulations, Section 3 the halo profiles and analytical models to fit them, while Section 4 briefly compares the simulations to the mass profiles of A2218 estimated by strong and weak gravitational lensing.

## 2. The simulations

I selected the nine most massive clusters from a cosmological simulation (White 1994, priv. comm.) of an Einstein-de Sitter universe with zero cosmological constant and scale free spectrum of density perturbations: $P(k) \propto k^n$, $n = -1$. The simulation is 150 Mpc on a side, for a dimensionless Hubble parameter $h = 0.5$. I resampled with higher resolution the initial conditions of each cluster, and re-run nine high resolution simulations. For the evolution I used the tree-SPH code described in Navarro and White (1993), without gas, i.e. as a pure $N$-body code. This code has individual and arbitrary time stepping (Groom and White 1996, in preparation). The normalization for all simulations is such that at the final time the rms matter density fluctuation in spheres of radius $r = 8h^{-1}$ Mpc is $\sigma_8 = 0.63$, in agreement with the observed abundance of local clusters (White et al. 1993).

The Virial mass of the clusters, enclosing an average overdensity $\delta\rho/\rho = 178$, ranges from $M_V = 5.2 \times 10^{14} M_\odot$ to about $3 \times 10^{15} M_\odot$, while their one dimensional rms velocity within the Virial radius ranges from 900 km s$^{-1}$ to 1350 km s$^{-1}$. The average number of particles within the Virial radius of each cluster is $N_V \simeq 20,000$. The gravitational softening $s$ imposed on small scales follows a cubic spline profile, and is kept fixed in physical coordinates. Its value is $s = 20 - 25$ kpc at the final time, depending on the simulation. The typical maximum number of timesteps per simulation is of order 20,000. All simulations were run on top-line workstations.

Profiles were defined by binning the particles in spherical shells centered on the cluster center, found iteratively as the the center of mass of a region of increasingly smaller size. The bins were equally spaced logarithmic intervals of width 0.1. Several tests were performed to establish what is the effect, on the resulting halo profiles, of varying the softening parameter $s$ of the simulation and the number of particles $N_V$ forming the final halo. These tests show that, for the softening value taken in the nine simulations of the sample, $s = 20 - 25$ kpc, the number of particle used, $N_V \approx 20,000$, is large enough to maintain the system collisionless even in its densest regions. They also show that the profiles are independent on numerical accuracy to better than 20% at all radii larger than the gravitational softening parameter $s$.

## 3. The profiles: Results and analytical fits

Here I will show the profiles of the nine halos and compare them to two analytic fits recently proposed in the literature: the Hernquist (1990) profile (HER) and the Navarro, Frenk and White (1995) profile (NFW). Both describe the density profiles as a two power-law curve, governed by a single *scale parameter* that sets the scale of transition between the two power laws. Their analytic expressions are: $\rho(r) = \rho_0/x(1+x)^2$ (NFW) and $\rho(r) = \rho_0/x(1+x)^3$ (HER); $x = r/r_0$ and $r_0$ is the scale parameter of the fit. Analogous expression for the circular velocity are easily derived from these. At small radii both profiles has the same asymptotic behaviour: $\rho(r) \propto r^{-1}$. At large radii the NFW fit is shallower, as $\rho(r) \propto r^{-3}$, while for HER it is $\rho(r) \propto r^{-4}$. In the range of interest, i.e. for $0.01 \le r/r_V \le 1$ (with $r_V$ the Virial radius), the Hernquist profile is the more curved of the two, because the difference between the two power law is larger. The density profiles of the nine dark matter halos were fitted with both analytic models in the range $[0.01 r_V, r_V]$, using standard Chi-square techniques. The profiles of each of the nine halos were averaged over the last four snapshots of each simulation, roughly corresponding to outputs between $z = 0.1$ and $z = 0$. This was done to obtain an *average* dynamical configuration for the halos, and to remove possible transient phenomena.

Figures 1 and 2 show the density and circular velocity profiles respectively, for these averaged configuration, and their best NFW and HER analytical fit. The thick solid curve is the simulated halo profile, the dotted curve is the best fitting NFW profile, the dashed curve is the best fitting HER profile. The fit has been made on the density profiles, therefore in some cases the best-fitting curve for the density does not correspond to the best-fitting curve for the circular velocity.

On average, the density profiles show a steeper logarithmic slope at large radii, an intermediate region where the slope bends towards the singular isothermal sphere value $-2$, and finally an inner region where the profile becomes gradually shallower. In some cases the intermediate region follows almost a power law; in others the change of slope is more gradual and relatively smooth. For the circular velocities, the general trend is a rising curve at small radii, followed by a maximum and then a decrease at large radii. The departures from the isothermal behaviour $v_c = constant$ are not due to limited resolution, and in particular, the decrease at large radii starts always within the Virial radius of the clusters.

Although the fits are fairly good, on average the density profiles of the simulated clusters look steeper than both analytical fits at small radii, and shallower than both at large radii. Part of the positive residuals at large radii may be due to the presence of merging subclumps in the profiles. The scatter (one $\sigma$ dispersion) in the residuals is smaller for the NWF profile, about 20% in density, and larger for the HER profile, about 30%. The circular velocity is fitted with smaller scatter, 8% and 13% for NWF and HER respectively. Considering the fact that the curves extend over two order of magnitude in radius, and over 4 order of magnitude in density, both fits can be considered quite good. The NFW model fits our simulations on average 50% better than the HER model, down to at least $r \simeq 0.01 r_V$. Since the simulated profiles are in general less curved than either the models, and are flatter at intermediate radii, the maximum circular velocity estimated by the fits is generally bigger than the true one, although the difference is small.

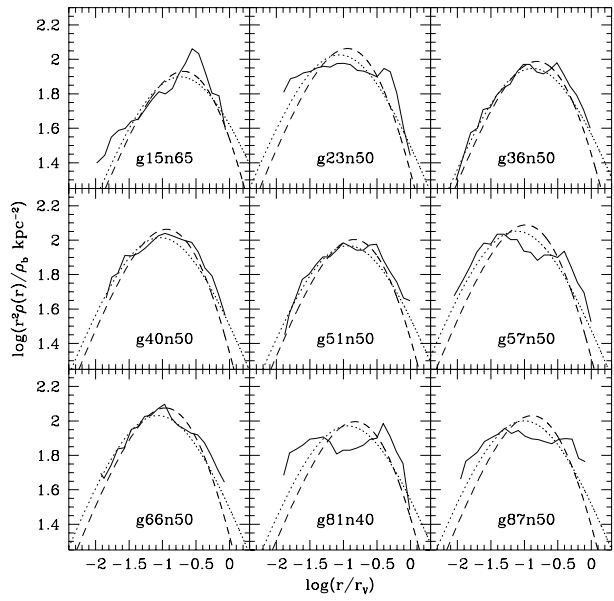

Figure 1.  NFW and HER fits to average density profiles.

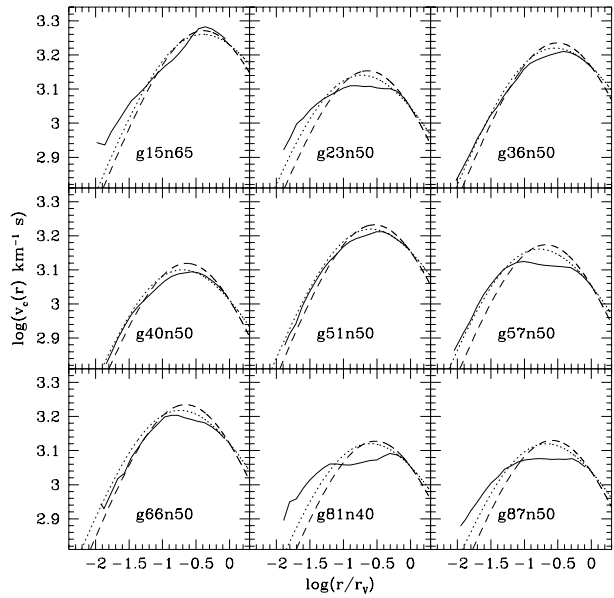

Figure 2.  NFW and HER fits to average circular velocity profiles.

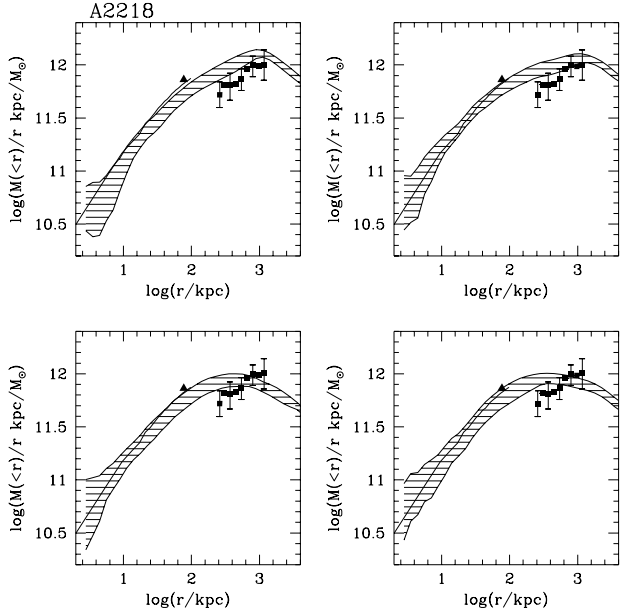

Figure 3. Mass profiles for A2218 plus four different projected halos from the simulations. See text for details.

## 4. Consequences on Cosmology: Gravitational Lensing

An interesting application of the present study is the comparison of the mass and density profiles of the simulations with those coming from observations of weak and strong gravitational lensing (GL) in rich clusters of galaxies. In particular, since the profile are resolved down to few tens of kpc from the center, these simulations can be compared to the mass estimates at very small radii provided by giant arcs. To this purpose, the simulated halos were considered at different times; for each output their particles were projected 100 times along random lines of sight (l.o.s.), and projected profiles were computed in equally spaced logarithmic bins, following the same procedure used for the radial profiles.

Figure 3 shows in each panel the mass profile measured for A2218. The solid curve is a mass model inferred by combined arc and arclets observations (Kneib et al. 1995). The triangle is an estimate from a giant arc (Miralda Escudè and Babul, 1995). The solid squares comes from a weak lensing model (Squires et al. 1995). The weak lensing measure is only a lower limit on the mass, and in fact it falls slightly lower than the other estimates. These observations are compared to four outputs of the simulations (indicated as shaded strips), two from each of the two most massive halos in the simulations. These halos have a velocity dispersion comparable to that estimated for A2218. The strips are centered on their mean projected mass, and enclose the ± one-$\sigma$ of the distribution of the 100 projections. As one can see, the simulated halos exhibit projected profiles fully consistent with the observed one. Very good agreement has also been found between the simulations and GL observations of other rich galaxy clusters.

**Acknowledgments.** I would like to thank Simon White for providing the results of his simulations, and for useful discussions and support. Thanks also to Julio Navarro for providing his tree-SPH code, and to Wendy Groom for providing the time integrator. The work presented here is part of a larger collaboration involving F. Bouchet (I.A.P.) and S. White (M.P.A.).